\documentclass[12pt]{article}

\usepackage{latexsym}

\def\be{\begin{equation}}
\def\ee{\end{equation}}
\def\ba{\begin{array}}
\def\ea{\end{array}}
\def\bea{\begin{eqnarray}}
\def\eea{\end{eqnarray}}
\def\bd{\begin{displaymath}}
\def\ed{\end{displaymath}}

\def\nn{\nonumber}
\def\qq{\quad\quad}

\newcommand{\ct}[1]{\cite{#1}}

\newcommand{\fr}[2]{{\frac{#1}{#2}}}
\newcommand{\re}[1]{(\ref{#1})}

\def\ra{\rightarrow}

\def\half{\frac{1}{2}}
\def\su{\sqrt{-u^2}}

\def\a{\alpha}
\def\b{\beta}
\def\g{\gamma}

\def\d{\delta}

\def\e{\epsilon}
\def\ve{\varepsilon}

\def\f{\phi}

\def\l{\lambda}

\def\m{\mu}
\def\n{\nu}

\def\g{\gamma}

\def\cL{{\cal L}}
\def\cH{{\cal H}}

\newcommand{\dr}{\raise.3ex\hbox{$\stackrel{\leftarrow}{\partial }$}{}}
\newcommand{\delr}{\raise.3ex\hbox{$\stackrel{\leftarrow}{\delta }$}{}}
\newfont{\twolineletters}{msbm10}
\newcommand{\Rrbar}{\mbox{\twolineletters R}}

\csname @addtoreset\endcsname{equation}{section}

\newtheorem{theorem}{Theorem}

\newtheorem{lemma}{Lemma}

\csname @addtoreset\endcsname{lemma}{section}

\newlength{\blength}
\newcommand{\proof}[1]{\vspace{-.05cm}
\begin{list}{\bf
{Proof:}}{\listparindent=\parindent\parsep=0pt
\labelwidth=0cm
\labelsep=\parindent
\addtolength{\labelsep}{-\blength}
\addtolength{\labelsep}{0.8cm}
\itemindent=-\blength
\addtolength{\itemindent}{\parindent}
\leftmargin=0.8cm}
\item 
#1~$\Box$
\end{list}
\vspace{.0cm}}

\begin{document}

\begin{titlepage}
\begin{raggedleft}
ULB--TH--01/07\\
KUL-TF-2001/10\\
{\tt hep-th/0104048}\\[.5cm]
\end{raggedleft}
\ \vskip 3.0mm
\begin{center}
{\bf \Large Deformations of duality-symmetric theories}
\vskip 5.3mm
{\bf Xavier Bekaert$^{a \,\clubsuit}$ and Sorin Cucu$^{b \, *}$}
\vskip 8.0mm
{$^a$ \em Physique Th\'eorique et Math\'ematique, Universit\'e Libre de Bruxelles,}\\
{\em Campus Plaine C.P. 231, B-1050 Bruxelles, Belgium} \\
{$^b$ \em Instituut voor Theoretische Fysica, Katholieke
Universiteit Leuven,}\\ {\em Celestijnenlaan 200D, B-3001 Leuven, Belgium}\\
\end{center}

\vskip 0.8cm

\begin{center}
{\bf Abstract}
\end{center}
We prove that a sum of free non-covariant duality-symmetric actions does not allow consistent, continuous and local
self-interactions that deform the gauge transformations. For instance, non-Abelian deformations are not allowed, even in 4 dimensions
where Yang-Mills type interactions of 1-forms are allowed in the non-manifestly duality-symmetric formulation.
This suggests that non-Abelian duality should require to leave the standard formalism of perturbative local field theories. 
The analyticity of self-interactions for a single duality-symmetric gauge field in four dimensions is also analyzed.
\vfill
\hrule width 5 cm

{\small
\noindent $^\clubsuit$ E-mail: xbekaert@ulb.ac.be \\
$^*$ E-mail: sorin.cucu@fys.kuleuven.ac.be}

\normalsize

\end{titlepage}
\newpage

\section{Introduction}

\par 
Duality symmetry has a rather long history, going back to the birth of Maxwell equations. As the field equations were invariant under the ``rotation" of the electric and magnetic fields, the physicists aimed to make this symmetry manifest in the action itself. After some first attemps in the seventies \cite{z:71,dt}, the topic has been more systematically addressed in the last decade in the view of its applications in supergravity and string theory. This efforts
led to quadratic but non-covariant versions \cite{dt,ss}, to quadratic and covariant actions but with an infinite number of auxiliary fields \cite{McClain:1990}, or to non-polynomial Lagrangians with manifest space-time symmetry \cite{pst}.

The electric-magnetic duality symmetry is deeply related to chiral forms, and can be elegantly reformulated in terms of a
self-duality condition on a complex field strength.
In order for the self-duality condition to be meaningful, the duality operation has to be an involution.
Because of the Minkowskian signature, the square of the Hodge dual $*$ can only be an involution in twice odd dimensions,
where chiral forms are well defined. In twice even dimensions, we have to take a complex field and redefine the duality
operator to be ${\rm i}*$ instead of the Hodge dual itself. 
The complexification of the fields is equivalent to the dualization of a pair of real gauge fields, as it is commonly done nowadays.

The main goal of this paper is to analyze the local consistent deformations of a system of Abelian $p$-form gauge fields
described by a sum of free duality-symmetric actions. A motivation for this topic is that a Yang-Mills extension of
duality-symmetric actions would be of interest in the attempt of generalizing Hodge duality to non-Abelian gauge groups. From the old analysis of \cite{dt,Gu} we know that mere Hodge duality cannot be consistently implemented for Yang-Mills theory.
But this does not prevent to try less ``trivial" generalizations of Abelian duality that should not be, a priori, too exotic.

Using the cohomological reformulation for the study of consistent deformations \cite{Barnich:1993}, we try to find if non-Abelian, local, continuous deformations of a sum of non-covariant gauge-fixed duality-symmetric actions are allowed or not. This will lead us to a no-go theorem suggesting that non-Abelian duality requires more unusual properties than what could be naively expected. 
In order to describe correctly non-Abelian duality, it seems mandatory to leave the standard formalism of perturbative local field theory.
Throughout this paper we deal with $U(1)$ vector gauge fields in four
dimensions but results analogous to the no-go theorem are also valid for higher rank forms in higher dimensions. 

A similar no-go theorem was found previously for chiral two-forms in six dimensions \cite{Bekaert:1999}.
In fact, the no-go theorem presented here is then not too surprising since duality-symmetric Maxwell theory in four dimensions
can be obtained by a dimensional reduction of a chiral two-form \cite{Verlinde} in six dimensions.

Self-interactions for a single duality-symmetric gauge field in four dimensions (i.e. sourceless non-linear electrodynamics) represent the second topic addressed in this paper. It can be traced back to the electric-magnetic duality symmetry of Born-Infeld action \cite{BI}, noticed by Schr\"odinger long time ago \cite{Schr}. After an analysis of Gaillard and Zumino \cite{Zumino} of duality invariance, Gibbons and Rasheed found the necessary and sufficient condition for
electric-magnetic duality invariance of non-linear electrodynamics \cite{GR}. The Lagrangian of the theory has to fulfill a non-linear partial differential equation of two variables, which has been re-obtained later from different approaches in \cite{PS,DS,XH}. Extending the analysis of \cite{PS}, we determine the necessary and sufficient set of conditions for the analyticity (in the weak field limit) of solutions to this differential equation.

The paper is structured as follows. Section \ref{s:dual} presents the duality-invariant formulations of the free Maxwell theory which constitute the starting point of this paper.
Section \ref{s:deform} gives the main results of this paper, that is the complete classification of all non-trivial consistent local deformations
of the non-covariant action presented in section \ref{s:dual}. 
After a brief review on the interpretation of consistent deformations as elements of the local BRST cohomology,
the long section \ref{s:cohomology} is devoted to the proof of the two theorems given in section \ref{s:deform}.
In the light of these theorems, the self-interactions for a single duality-symmetric gauge field in four dimensions are analyzed in section \ref{s:self}.
There we re-derive in the PST approach, the equation enforcing duality invariance and we also present an extension of previous results about the analyticity of solutions to this equation.
In section \ref{s:concl}, we conclude with a brief summary of the main results presented in this paper.

The paper is self-contained and contains all the technical details necessary to prove the main results of this paper.
To make the paper more legible the technical lemmas, their proofs and some notations have been placed in the appendices.

\section{Duality-symmetric free action}\label{s:dual}

As we aim to discuss possible deformations for duality-symmetric theories it will be instructive to recall first the free model in both its covariant and non-covariant formulation. The starting point is given by the PST action that besides the electric-magnetic duality exhibits also manifest Lorentz invariance. After fixing the notation, we show that a suitable gauge choice leads to the non-covariant action.    

The action proposed by PST \cite{pst} for the description of a duality-symmetric vector field (two potential formulation) is 
\begin{eqnarray}
S_0&=& \frac{1}{2}\,\int d^4x \, h^\a_m \tilde \cH^{\a m} \, 
\label{e:pst}
\end{eqnarray}
where the $m,n,\dots\,$ stand for Lorentz indices in $4$ dimensional space-time with a flat metric $(-,+,+,+)$. The Lagrangian contains two gauge potentials $(A_m^\a)_{\a=1,2}$ (related by a duality relation) together with the gradient $u_m=\partial_m a$ of an auxiliary field $a$. Here we employed the following notation   
\begin{eqnarray}
u^2=u^mu_m \,, &  & v_m=\fr{u_m}{\su}\,,\\
F^\a_{mn}=2\partial_{[m}A^\a_{n]}\,, & & F^{*\a}_{mn}=\half \,\e_{mnpq}F^{\a pq} \,,\\ 
\cH^\a_{m} = \cL^{\b\a}v^n F^\b_{nm} \,, &  & \tilde \cH^\a_{m} = v^n F^{*\a}_{nm}\,,\\  
 & h^{\a}_{m}= \cH^\a_{m} -  \tilde \cH^\a_{m} \,,&  
\label{e:not}
\end{eqnarray}
with $\cL^{12} = - \cL^{21} = 1$. The fact that $a$ is auxiliary can be understood directly as its field equation follows as a consequence of the equations of motion for $A_m^\a$, which read
\begin{eqnarray}
\e^{mnpq}\partial_n(v_p h^\a_q)=0\,.
\label{e:eomA}
\end{eqnarray}

It is not difficult to check that the free action $S_0$ is invariant under the gauge transformations
\begin{eqnarray}
\d_IA_m^\a=\partial_m\varphi^\a\,, &\,\d_I a=0 \,,\label{e:ginv1}\\
\d_{II}A_m^\a= \cL^{\b\a} h^{\b}_{m}\fr{\f}{\sqrt{-u^2}}\,, &\,\d_{II}a=\f \,,\label{e:ginv2}\\
\d_{III}A_m^\a=u_m\ve^\a\,, &\, \d_{III}a=0 \,.\label{e:ginv3}
\end{eqnarray}
One can use a suitable gauge-fixing \ct{pst} for (\ref{e:ginv3}) in order to express the solution of (\ref{e:eomA}) in the form of a self-duality condition
\begin{equation}
\cH^\a_{m} - \tilde \cH^\a_{m} = 0\,.\label{e:duality}
\end{equation}
Moreover, the invariance \re{e:ginv2} states the gauge character of the scalar $a$. These last two remarks show that the physical content of the free theory reduces to only one gauge vector (the other one can be written in terms of this one by means of the duality relation \re{e:duality}, while $a$ is pure gauge)\footnote{These fields were needed only to implement both self-duality and Lorentz invariance at the level of the action.}. 

Pasti, Sorokin and Tonin have shown \cite{pst} that their model is in fact classically equivalent to the non-covariant
two-potential action \cite{dt,ss} describing also the dynamics of a single Maxwell field
\begin{equation}\label{e:SchwarzSen}
S=-{1\over 2}\int d^4x(B^{i\alpha}{\mathcal L}^{\alpha\beta}E^\beta_i
+B^{i\alpha}B_{i}^\alpha),
\end{equation}
where
\begin{equation}\label{eb}
E^\alpha_i=F^\alpha_{0i}=\partial_0A^\alpha_i-\partial_iA^\alpha_0,
\qquad
B^{i\alpha}={1\over 2}\varepsilon^{ijk}F^\alpha_{jk}=
\varepsilon^{ijk}\partial_jA^\alpha_k,
\end{equation}
and $i,j,k = 1,2,3$ are spatial indices. Written in this way the action is only {\it manifestly symmetric} under rotations but not under Lorentz boosts (even it is Lorentz invariant). This invariance descends from the gauge transformations \re{e:ginv2}. Moreover it possesses the usual gauge invariance
\begin{equation}
\d A_i^\a=\partial_i\varphi^\a\,, 
\end{equation}
residual of (\ref{e:ginv1}). A simple gauge choice $u^m=(1,0,0,0)$ in \re{e:pst} leads directly to (\ref{e:SchwarzSen}). 
Historically, the PST action (\ref{e:pst}) has been obtained precisely by ``covariantizing" \re{e:SchwarzSen}. The price paid for keeping the Lorentz covariance as a symmetry for the action $S_0$ is given by an extra gauge transformation \re{e:ginv2} with respect to the non-covariant formulation \re{e:SchwarzSen}. It is precisely this symmetry that will be deformed when looking at the interacting theory. 

Using a ``formal" path integral quantization ("formal" in the sense that the possible UV divergences due to the non-Gaussian character of the integral were not considered) it has been proved in \cite{Sorin} that the  equivalence mentioned above remains valid also at the quantum level. The absence of anomalies and non-trivial counterterms has been shown recently in \cite{Piguet}.

\section{Consistent deformations of the non-covariant action}\label{s:deform}

We want to analyze the consistent deformations of a system of $N$ free Abelian $1$-forms (on-shell) described by the sum of $N$
non-covariant actions, manifestly invariant under duality. We are looking for a Yang-Mills extension of this system. 

We know that for a set of $N$ Abelian $1$-forms described by a sum of Maxwell actions, the Yang-Mills theory is the unique
Lorentz invariant consistent deformation of the model deforming the gauge symmetry (see e.g. \cite{uniqueness}).
But non-Abelian duality seems to intrinsically involve non-local or non-perturbative features. This is related to the
non-local transformations that relates the theory in Hamiltonian formalism to its (conjectured) duality-symmetric
formulation. 
For example, the standard Hamiltonian procedure for usual Maxwell theory gives $\pi_i=F_{0i}={\dot A}_i-\partial_iA_0$.
Hence the potential $A_0$ can be expressed non-locally in terms of the other variables (e.g. by a curve integral).
Thus, the Yang-Mills cubic vertex containing explicitly $A_0$ would be non-local if we express it in terms of the other variables
to make the link with the duality-symmetric formulation. In any case (even in temporal gauge), non-locality will arise from
solving perturbatively
the Gauss law $D_i\pi^i=0$ for $\pi^i$, and inserting its non-local expression in the cubic vertex.
Hence it could be expected that such a non-Abelian, local, continuous deformation of a sum of manifestly invariant actions
under duality is not possible at all.
And indeed, this is what we found.

The main theorem presented here is
\begin{theorem}\label{t:classification}
All consistent, continuous, local deformations of a system of free Abelian vector fields ($A=1,\cdots,N$)
described by a sum of $N$ free, duality-symmetric, non-covariant actions as the coupling constant goes to zero, are only of two
types:
\begin{itemize}
\item[I.]{those which are strictly invariant under the original gauge transformations; they are polynomials in curvatures
and their partial derivatives, i.e.
of the form 
\begin{equation}\int d^4x\,f(\partial_{i_1\dots i_k}F^{\a A}_{ij})\,.\end{equation}} 
\item[II.]{those which are invariant only up to a boundary term;
they are linear combinations of Chern-Simons like terms, i.e. 
\begin{equation}\int d^4x\,\lambda_{\a\b\,ab\,AB}\,(\partial_0)^aA_i^{\a A}(\partial_0)^bB^{\b B\, i} \,,\end{equation}
where $\lambda_{ab\,\a\b\,AB}$ are constants such that 
$\lambda_{ba\,\b\a\,BA}=\lambda_{ab\,\a\b\,AB}\,$.}
\end{itemize} 
\end{theorem}
For instance, the Hamiltonian of the free theory is a term of type I, as well as the Hamiltonian for Born-Infeld theory. 
The kinetic term is of type II.

As a corollary, the following no-go theorem holds
\begin{theorem}\label{t:nogo}
No consistent, local interactions of a set of free Abelian vector fields can deform the Abelian gauge transformations
if the local deformed action (free action + interaction terms) continuously reduces to a sum of free, duality-symmetric,
non-covariant actions in the zero limit for the coupling constant. 
\end{theorem}

We want to stress that analogous theorem holds for any duality-symmetric theories in twice even dimensions. 
We will only give here the proof for the four dimensional case\footnote{Notice that for higher dimensions the proof follows the same procedure but is much more straightforward because of
the non-matching of some degree. This is not too surprising because in general forms of higher rank $p$ are known to have a high degree of rigidity \cite{BernardK}.}.
Next, we present an useful algebraic reformulation of the consistent deformation problem, which will allow us to use the
powerful technique of cohomological analysis in proving these theorems.

\section{Constructing consistent deformations as a cohomological problem}\label{s:cohomology}

A deformation of a theory is called {\em{consistent}} if the deformed theory possesses the same number of independent gauge symmetries, reducibility identities, etc, as the system we started with. In other words, the number of physical degrees of freedom is unchanged.
The problem of deforming an action consistently is in fact equivalent to the problem of deforming the solution $S$ of the
master equation $(S,S)=0$ into a solution $S'$ of
the deformed master equation $(S',S')=0$ (Basically, this is due to the fact the master equation encodes
completely the gauge structure of a theory). 
We can treat this last problem perturbatively, 
that is, assume that $S'$ can be expressed as a power series in a dimensionless coupling constant $g$, 
the zeroth-order term in the expansion of $S'$ being equal to the solution of the free theory $S$. 
Then we try to construct the deformations order by order in $g$.
It can be shown that the first-order non-trivial deformations of $S$ are elements of the BRST cohomology group in vanishing ghost number (denoted here as {\it gh} number) : 
$H_0(s)$ \cite{Barnich:1993}. This is nothing else than an elegant reformulation of the well-known Noether method.

If we now require locality, the non-trivial deformations of the theory will be even more constrained, because
in that case we restrict the deformation to be a local functional, i.e. $S'=\int \alpha^n$, where the integrand
$\alpha^n$ is a {\em local $n$-form} in $n$ dimensions, that is a differential $n$-form with local functions as
coefficients.
{\em Local functions} depend polynomially on the fields
(including the ghosts and the antifields) and their derivatives up to a finite order
(in such a way that we work with functions over a finite-dimensional vectorial space, the so-called jet spaces).
The non-trivial deformations $\a^n$ are elements of the local BRST cohomology group in {\it gh} number zero
\cite{Barnich:1993}, that is $H^n_0(s|d)$ 
where $d$ is the spacetime differential.

\subsection{Field content, BRST operator}

We apply the Batalin-Vilkovisky formalism \cite{bv} to a sum of $N$ non-covariant actions (\ref{e:SchwarzSen}) in the
temporal gauge ($A^{\a}_0=0$) and find that
\begin{equation}
S= \sum_{A=1}^N  \int d^4 x\,\left[\fr12 (\cL^{\a\b}\dot{A}^{\a A}_i -\delta^{\a\b}B^{\a A}_i)B^{\b A\,i}+\int d^4 x\,A^{\a
A*}_i\partial^iC^{\a A}\right]
\end{equation}
is a minimal solution to the master equation.
The field content of the theory consists of the two potentials $A^{\a A}_i$, their ghosts $C^{\a A}$ 
and the associated antifields $A^{\a A *}_i$ and $C^{\a A *}$. Their respective statistics, ghost number, antighost number are listed in Table \ref{ta:}. 
Notice that $gh\equiv puregh - antigh$.

Because the theory is Abelian, the BRST operator of this theory is simply the sum of the Koszul-Tate differential $\delta$
and the differential $\gamma$. They act on the fields and antifields in the following way
\begin{eqnarray}
\delta A^{\a A}_i&=&\delta C^{\a A}=0,\\
\delta A^{\a A*}_i&=& \partial^jF_{ij}^{\a A}+\fr12\cL^{\a\b}\epsilon_{ijk}\dot{F}^{\b A\,jk},\\
\delta C^{\a A*}&=& - \partial^i A^{\a A*}_i,\\
\gamma A^{\a A}_{i}&=&\partial_i C^{\a A},\\
\gamma C^{\a A}_i&=&0,\\
\gamma A^{\a A*}_i&=&\gamma C^{\a A*}=0.
\end{eqnarray}
Furthermore, $\delta$ and $\gamma$ commute with the partial derivatives $\partial_m$ and act trivially on the coordinates
$x^m$, thus their action on all the generators of the algebra of local forms can be easily determined using the Leibnitz rule.
Since $S$ is solution to the master equation, the BRST operator is nilpotent $s^2=0$ and one verifies that this implies $\delta^2=\delta\gamma+\gamma\delta=\gamma^2=0$.

In the sequel we will work in the algebra of local forms that is generated by the previous fields (and antifields) and all
their spacetime derivatives up to a finite order, as well as by the coordinates $x^m$ and their differentials $dx^m$.
\begin{table}[h]
\centerline{
\begin{tabular}{|c|c|c|c|c|}
\hline
 & statistics & antigh & puregh & gh\\
\hline
$A_i^{\a A}$ & + & 0 & 0 & 0\\ \hline
$C^{\a A}$ & - & 0 & 1 & 1\\ \hline
$A_i^{\a A *}$ & - & 1 & 0 & -1 \\ \hline
$C^{\a A *}$ & + & 2 & 0 & -2 \\ \hline
$s$ & - & - & - & 1 \\ \hline
$\delta$ & - & -1 & 0 & 1 \\ \hline
$\gamma$ & - & 0 & 1 & 1 \\ \hline
\end{tabular}
}
\caption{Respective statistics, ghost number, antighost number of the variables and the differential operators}
\label{ta:}
\end{table}

\subsection{Local BRST cohomology in ghost number zero}\label{ss:H0}

To compute the local BRST cohomology we will follow the standard procedure \cite{Barnich:1995} (see also the recent review
\cite{Barnich:2000}). 
The computation is very similar to the one developed in \cite{Bekaert:1999} for the consistent deformations of a system of
free chiral $p$-forms, 
which are relevant for the study of coinciding M5-branes. 

Let $\alpha^4$ be a local 4-form of vanishing {\it gh} number. 
To be an element of $H^4_0(s|d)$ it has to be a non-trivial BRST cocycle modulo $d$.
In other words, it must satisfy the Wess-Zumino consistency condition
\begin{equation}
s\alpha^4 + d\beta^3 = 0 \,,
\label{e:WZ}
\end{equation}
where $\beta^3$ is an arbitrary local 3-form of {\it gh} number 1. 
The cocycles $\alpha^4$ of $H^4_0(s|d)$ are defined up to the following equivalence relation
\begin{equation}
\alpha^4 \sim  \alpha^4 + s\rho^4 + d\sigma^3
\end{equation}
with $\rho^4$ and $\sigma^3$ arbitrary forms of {\it gh} number $-1$ and $0$ respectively.
A trivial element is an exact cocycle of $s$ modulo $d$ and it can always be identified to zero.

It can be proved that if the component of {\it antigh} number zero of $\alpha^4$ 
(that is the first-order deformation of the action)
is local and if $\alpha^4$ satisfies the W-Z consistency condition, 
then we can assume in complete generality that each term in its {\it antigh} number expansion is local 
and the sum stops at a finite order (see lemma \ref{l:expansion} in the appendix \ref{a:locality}).
Therefore, it is enough to postulate the locality of first-order deformation of the \textit{action} (as it is done in
theorems \ref{t:classification} and \ref{t:nogo})
in order to enforce the locality of first-order deformation of the \textit{solution} to the master equation 
(which contains also the gauge transformations, etc).

In conclusion, we can expand $\a^4$ according to the {\it antigh} number as
\begin{equation}
\alpha^4 = \alpha^4_0 + \alpha^4_1 + \dots + \alpha^4_k \,,
\end{equation}
where $\alpha^4_i$ stands for the {\it antigh} number $i$ component of $\alpha^4$. Then, the component of {\it antigh}
number $k$ of (\ref{e:WZ}) is
\begin{equation}
\gamma\alpha^4_k + d\beta^3_k = 0.
\label{e:gammamodd}
\end{equation}
This equation tells us that $\a^4_k$ is a cocycle of $\gamma$ modulo $d$. Furthermore, any trivial part in $\a^4_k$ can be
removed by an appropriate redefinition,
e.g. if $\a^4_k=\gamma \rho^4_k + d\sigma^3_k$, the redefinition would be
\begin{equation}
\alpha^4 \rightarrow \alpha^{4'} =  \alpha^4 - s\rho^4_k - d\sigma^3_k.
\end{equation}
Hence $\a^4_k$ belongs to $H^4_k(\g |d)$. 

The case $k=0$ corresponds to the case where the deformation of the action is invariant under the original gauge
transformations 
up to a boundary term, so the gauge transformations need not be deformed. 
In that case, (\ref{e:WZ}) completely reduces to (\ref{e:gammamodd}),
and so we are interested in the cohomology of $\gamma$ modulo $d$ in form degree 4, vanishing ghost and {\it antigh}
numbers: $H^4_{0,\,0}(\gamma|d)$.
This group is given by lemma \ref{l:gammamodd0}.

We should also notice that, more precisely, the first-order {\em non-trivial} deformations that do not deform the gauge
transformations are elements of $H^4_0(\gamma|d,H_0(\delta))$. Indeed, if a cocycle is $\delta$ exact, then it corresponds to a trivial
first-order deformation corresponding to a mere redefintion of the fields. 
For example, the elements of type I simply belong to $H_{0,0}(\gamma)$.
But if they depend on the $\bar{F}^{\a A}_{ij}$'s and their derivatives (see lemma \ref{l:gamma} for their definition),
then they define trivial first-order deformations.
Therefore, we should only keep the terms involving no time derivatives of the curvatures: 
$f(\partial_{i_1\dots i_k}F^{\a A}_{ij})\,d^4x$. This provides the two types of deformations given in theorem
\ref{t:classification}.

Now, we have to show that all local first-order interactions that deform non-trivially the gauge transformations are either inconsistent or trivial.
Let us suppose that the gauge transformations are deformed at first-order, i.e. $k>0$.
In that case, it can be proved that, without loss of generality, we can assume $d\beta^3_k=0$ in (\ref{e:gammamodd})
(see lemma \ref{l:gammamodd'}). Therefore, $\alpha^4_k$ belongs to the cohomology group $H(\gamma)$.
Hence, 
\begin{equation}\label{e:invariantk}
\alpha^4_k=dx^0\sum_I \tilde P^3_I(\chi)\omega^I + \gamma \m^4_k,
\end{equation}
where the $\tilde P^3_I(\chi)$ denote invariant local spatial forms of degree 3 (see App. \ref{a:gamma}). 
The $\gamma$-exact component of $\a^4_k$ can be eliminated from this component by redefining $\a^4$ as being equal to
$\a^{4'}:=\a^4-s\m^4_k$.
Now, we can go one stage lower.

The component of {\it antigh} number $k-1$ of (\ref{e:WZ}) is
\begin{equation}
\delta\alpha^4_k + \gamma\alpha^4_{k-1} + d\beta^3_{k-1} = 0.
\label{e:k-1}
\end{equation}
Acting with $\gamma$ on (\ref{e:k-1}), and using equation (\ref{e:invariantk}) together with the algebraic Poincar\'e lemma
(see Appendix \ref{a:Poincar}),
we find that 
\begin{equation}
\gamma\beta^3_{k-1} + d\rho^2_{k-1} = 0.
\end{equation}
It is possible to show that in such a case, even for vanishing $antigh$ number ($k=1$), we can assume without loss of
generality that $d\rho^2_{k-1}=0$ (see lemma \ref{l:gammamodd''}).
Hence, $\beta^3_{k-1}$ is a cocycle of $\gamma$.
Using this, and analysing carefully the equation (\ref{e:k-1}), it can be shown that by an allowed redefinition of
$\alpha^4$ and $\beta^3$, 
we can take $\beta^3_{k-1}=dx^0\sum_I \tilde Q^2_I(\chi)\omega^I$ (see \cite{Bekaert:1999}).
So (\ref{e:k-1}) becomes of the form
\begin{equation}
dx^0\sum_I\left(\delta \tilde P^3_I(\chi)+\tilde{d}\tilde Q^2_I(\chi)\right)\omega^I =\gamma\alpha'^4_{k-1},
\end{equation}
where $\tilde{d}=dx^i\partial_i$ is the spatial differential.
This implies 
\begin{equation}\delta \tilde P^3_I(\chi)+\tilde{d}\tilde Q^2_I(\chi)=0\,\end{equation} 
from \re{glll}. We can always remove $\delta$ trivial modulo $\tilde{d}$ cocycles from $\tilde P^3_I(\chi)$ from $\a^4$
by an appropriate redefinition of $\a^4$ that does not modify the condition $\gamma \a^4_k=0$.
Thus, $\tilde P^3_I(\chi)$ is a non-trivial element of the invariant
cohomology of $\delta$ modulo $\tilde{d}$: $H_k^{3,inv}(\delta|\tilde{d})\equiv H_k^3(\delta|\tilde{d},{\mathcal I}^*)$. From the lemmas \ref{l:delta1}-\ref{l:lastl} we learn that $\tilde P^3_I(\chi)$ is non-trivial only in {\it antigh}
number $k=2$. In that case, it corresponds to a deformation of the gauge algebra and is a candidate for a Yang-Mills type deformation.

In the next section we will prove that it is not possible to construct a local deformation of the action (in other words, there is an \textit{obstruction} to the descent of $\a^4_2$)
corresponding to this deformation of the gauge algebra. This ends the computation of the local BRST cohomology in {\it gh} number zero.

\subsection{Obstruction to a Yang-Mills type deformation}

Let $\a^4_2$ be determined by the following generic non-trivial element of $H_2^{3,inv}(\delta|\tilde{d})$ (see lemma \ref{l:delta1})
\begin{equation}\label{e:a2}
\a^4_2=\fr12 f_{abc\,\a\b\gamma\,ABC}\, (\partial_0)^aC^{\a A*} \cdot\,(\partial_0)^bC^{\b B}\cdot\,
(\partial_0)^cC^{\gamma C}\,d^4x \,. 
\end{equation}
There is an implicit sum on all repeated indices. By construction $f_{bc\,\a\b\gamma\,ABC}$ are constants satisfying 
\begin{equation}\label{e:antis}
f_{abc\,\a\b\gamma\,ABC} = - f_{acb\,\a\gamma\b\,ACB}\,,
\end{equation}
that is, they are antisymmetric with respect to the exchange of the last two indices in each set of indices.
This term can be interpreted as a candidate for a first-order deformation of the Abelian gauge algebra (with $f_{abc\,\a\b\gamma\,ABC}$ the structure constants).
We want to prove that there is no local $\a^4_0$ corresponding to the $\a^4_2$ given by (\ref{e:a2}).

We have to try to find a corresponding first-order deformation
of the gauge transformation, $\a^4_1$, by solving the equation
\begin{equation}
\delta \a^4_2 + \gamma \a^4_1 + d\b^3_1 = 0\,.
\end{equation}
There is no obstruction here and the answer is
\begin{equation}
\a^4_1= f_{abc\,\a\b\gamma\,ABC}\, (\partial_0)^aA^{\a A*\,i}\cdot\, (\partial_0)^bA_i^{\b B}\cdot\,(\partial_0)^cC^{\gamma
C}d^4x + c^4_1\,,
\end{equation}
where $c^4_1$ is a cocycle of $\gamma$ modulo $d$. 
Hence, $c^4_1$ has to correspond to the last term in the {\it antigh} number expansion of a cocycle $c^4$ of $s$ modulo
$d$: $c^4=c^4_0+c^4_1$, $sc^4+de^3=0$. 
This expansion stops at {\it antigh} number $1$ and then $c^4$ has to be trivial. So $c^4_1$ will be neglected in what
follows.

The last step is to find a corresponding local function $\a^4_0$ solution of
\begin{equation}\label{e:a0}
\delta \a^4_1 + \gamma \a^4_0 + d\b^3_0 = 0\,,
\end{equation}
which would be the first-order deformation of the action.
But we now show that there is an obstruction to this descent.
Firstly, we rewrite (\ref{e:a0}) explicitly
\begin{eqnarray}\label{e:a0'}
&&\fr12 f_{abc\,\a\b\gamma\,ABC}\left[ (\partial_0)^aF^{\a A\,ij} (\partial_0)^bF_{ij}^{\b B} +
\cL^{\a\d}\epsilon^{ijk}(\partial_0)^{a+1}F^{\d A}_{ij}(\partial_0)^bA_k^{\b B} \right]\times \nn\\
 && \times (\partial_0)^cC^{\gamma
C}d^4x = \gamma \mu_0^4 + d\nu_0^3 \,.
\end{eqnarray}
Let us decompose $\nu^3_0$ as $\nu^3_0=\tilde{\nu}_0^3+dx^0\tilde{\rho}_0^2$, in such a way that $d\nu_0^3$ takes the form
$dx^0(\partial_0\tilde{\nu}_0^3-\tilde{d}\tilde{\rho}_0^2)$.
We introduce the operator $N_{\tilde{\partial}}$ counting the number of spatial derivatives of fields. In particular,
we have
\begin{equation}
N_{\tilde{\partial}}(\gamma)=N_{\tilde{\partial}}(\tilde{d})=N_{\tilde{\partial}}(\partial_0)+1=1\,.
\end{equation}
The fact that $\gamma$, $\partial_0$ and $\tilde{d}$ are homogeneous in this degree implies
that the $F^2$ term and the Chern-Simons like term in the equation (\ref{e:a0'}) have to be $\gamma$ modulo $d$ exact
independently
because they have distinct $N_{\tilde{\partial}}$-degree. 
At  this point we have to ask ourselves what are the constraints imposed on the coefficients $f$ by the triviality of these
two terms.

The triviality of the $F^2$ term is encoded in
\begin{equation}
\sigma^4(I):=\fr12 f_{abc\,\a\b\gamma\,ABC}(\partial_0)^aF^{\a A\,ij}\cdot\, (\partial_0)^bF_{ij}^{\b B}\cdot\,
(\partial_0)^cC^{\gamma C}\,d^4x=\gamma a^4 + db^3\,.
\end{equation}
It implies $\gamma b^3 + dc^2=0$. From lemma \ref{l:gammamodd''} we know that we can assume
$b^3=\left(\tilde{b}_I^3(I)+dx^0\tilde{e}_I^2(I)\right)\omega^I$, which implies
\begin{equation}\label{e:F2}
\sigma^4(I)=dx^0[\partial_0\left(\tilde{b}_I^3(I)\omega^I\right)-{\tilde d}\tilde{e}_I^2(I)\cdot\omega^I] + \g a^{4'}\,.
\end{equation}
Now, (\ref{e:F2}) implies from (\ref{glll}) that $\g a^{4'}=0$ because we can take a basis of ${\mathcal C}$ such that the time derivative $\partial_0\omega^I$ are elements of the basis (see App. \ref{a:gamma} and \cite{Bekaert:1999}). 
Let $\tilde{N}$ be the degree counting the number of spatial derivatives of the curvatures $(\partial_0)^aF^{\a A\,ij}$.
Decomposing (\ref{e:F2}) according to the degree $\tilde{N}$ gives in vanishing degree
\begin{equation}
\sigma^4(I)=dx^0[\partial_0\left(\tilde{b}_{I,0}^3(I)\omega^I\right)\,,
\end{equation}
where $\tilde{b}_{I,k}^3(I)$ is the component of degree $\tilde{N}$ equal to $k$.
Indeed, $\sigma^4(I)$ is of degree zero and $\tilde{d}$ acting in the space ${\mathcal I}$ increases by one the number of spatial derivatives of curvatures.
It can be proved that any $(dx^0\partial_0)$-trivial part of $\sigma^4(I)$ corresponds to a $(dx^0\partial_0)$-trivial part of $\a^4_2$ (the inverse is trivial because
$\partial_0$ commutes with the operators $\delta_1$ and $\delta$). 
But a $(dx^0\partial_0)$-trivial part of $\a^4_2$ can be eliminated by a $d$-trivial redefinition of $\a^4_2$ since the very beginning,
hence we can assume that $\tilde{b}_I^3(I)$ and thus also $\sigma^4(I)$ vanish. This implies the following relation the coefficients $f$
\begin{equation}\label{e:-}
f_{abc\,\a\b\gamma \,ABC} = - f_{bac\,\b\a\gamma\, BAC}\,.
\end{equation}

The triviality of the Chern-Simons like term can be rewritten as
\begin{equation}\label{e:CS}
dx^0\,f_{abc\,\a\b\gamma\,ABC}\cL^{\a\d}(\partial_0)^{a+1}F^{\d A}\cdot\,(\partial_0)^bA^{\b B}\cdot\,
(\partial_0)^cC^{\gamma C}=\gamma a^{4'} + db^{3'}
\end{equation}
where $A^{\a A}:=A^{\a A}_idx^i$ and $F^{\a A}:=\tilde{d}A^{\a A}$.
To analyse the consequences of (\ref{e:CS}), we introduce the \textit{universal algebra} (see
\cite{Barnich:1995,Barnich:2000,knaepen} and references therein) generated
by $A^{\a A}$, $F^{\a A}$, $C^{\a A}$ and $\tilde{d}C^{\a A}$ with all their time derivatives up to a finite order,
\textit{without imposing any restriction on the maximally allowed form degree}. The universal algebra is of major
importance in the computation of $H(\gamma|d)$. One of the key properties of that space is that the Poincar\'e lemmas \ref{l:Poincalg}
and \ref{l:invP} remain valid also in form degree $\geq 4$.
Consistency imposes the equation
\begin{equation}\label{e:CS5}
dx^0\,f_{abc\,\a\b\gamma\,ABC}\cL^{\a\d}(\partial_0)^{a+1}F^{\d A}\cdot\,(\partial_0)^bF^{\b B}\cdot\,
(\partial_0)^cC^{\gamma C}=\gamma m^{5}\,,
\end{equation}
derived by applying the operator $d$ to (\ref{e:CS}). 
But this equation (\ref{e:CS5}) possesses a solution if and only if the $f$-coefficients satisfy
\begin{equation}\label{e:+}
f_{abc\,\a\b\gamma\,ABC} = \cL^{\rho\b}\cL^{\sigma\a} f_{(b-1)(a+1)c\,\rho\sigma\gamma\,BAC}\,,
\end{equation}
in such a way that each side of (\ref{e:CS5}) vanish.

Now we remark that the symmetry properties (\ref{e:-}) and (\ref{e:+}) on the coefficients imply that they all have to vanish.
Indeed, it is clear that all coefficients $f_{abc\,\a\b\gamma\,ABC}$ with $b=0$ must vanish
due to (\ref{e:+}). But, using alternatively (\ref{e:-}) and (\ref{e:+}) one can show that all the coefficients
$f_{abc\,\a\b\gamma\,ABC}$ are related linearly to the coefficents for which $b=0$. Therefore they all vanish.
This achieves the demonstration of the no-go theorem \ref{t:nogo}. 

In conclusion, all the consistent interaction are given by theorem \ref{t:classification}.
We can remark that our computation shows explicitly that the local BRST cohomologies of the usual (not manifestly invariant under duality) Maxwell theory  
and of its two-potential formulation (manifestly duality symmetric) are not isomorphic.

\section{Self-interactions of a single gauge vector}\label{s:self}

In this section we want to generalize the PST action describing a single {\it free} Maxwell field to an interacting theory with only one gauge vector (on-shell). The interacting model should of course maintain the Lorentz covariance and should lead to a deformed self-duality condition. Firstly, we review in Subsec. \ref{ss:C-Heq} how this has been achieved in the non-covariant approach, then we extend it in Subsec. \ref{ss:defPST} for the covariant case, and finally we look for the solutions of the Courant-Hilbert equation.

\subsection{Courant-Hilbert equation} \label{ss:C-Heq}

We want to introduce self-couplings for \re{e:SchwarzSen}.
In \cite{DS} it was proposed to attack this problem using the Hamiltonian formulation, which is appropriate to deal with
first-order actions.
Let us review their approach, trying to justify completely the ansatz in the light of the results obtained in the previous sections.

From the very beginning we state three basic requirements made on the model in this approach: (i)  The deformation of
\re{e:SchwarzSen} remains a first-order action, (ii) manifest rotation invariance, (iii)  manifest duality-symmetry.
The first assumption comes from the fact that we work in Hamiltonian formalism. 
The second and third requirements simply extend the properties of the free model.

The first requirement combined with theorem \ref{t:classification} and the fact that we deal with a single
on-shell vector gauge field ($N=1$) eliminates deformations of type $II$ besides the one without any derivative. 
Thus, the non-linear action generalizing \re{e:SchwarzSen} is 
\begin{equation}
S=-{1\over 2}\int d^4x\,B^{i\alpha}{\mathcal L}^{\alpha\beta}E^\beta_i - \int dx^0 H,
\end{equation}
where
\begin{equation}
H = \int d^3x \left({\mathcal H}(\partial_{i_1\dots i_{k-1}}B^{\a}_{i_k})+\chi A_i^\a B^{i\a}\right)
\end{equation}
stands for the Hamiltonian of the model. $\chi$ is the constant factor of the Chern-Simons like term. 
If we now restrict ourself to (iv) Hamiltonian densities that do not depend explicitly on the derivatives of the magnetic
field\footnote{In other words, we assume a slowly varying fieldstrenght.}: ${\mathcal H}={\mathcal H}(B^{\a A}_i)$,
we can deduce from (ii) and (iii) that the Hamiltonian density ${\mathcal H}$ only depends on two independent space scalars that are
manifestly duality invariant
\begin{equation}
y_1 = \frac12 B^{\a}_iB^{\a i}\,, \; \;
y_2 = \frac{1}{4} B^{\a}_iB^{\b i}B^{\a}_jB^{\b j}\,.
\end{equation}
We set ${\mathcal H}= f(y_1,y_2)$.

Now, all the symmetries are manifest except Lorentz symmetry. With the help of tensor calculus, it is rather easy to
construct interactions that preserve Lorentz invariance. But there is an alternative way to control Lorentz invariance. It is through
the commutation relations of the energy-momentum tensor components: Dirac and Schwinger gave a sufficient condition for a
manifestly rotation and translation invariant theory (in space) to be also Lorentz-invariant. The condition is necessary when one turns to gravitation. 
The Dirac-Schwinger criterion yields in our case $\chi=0$ and a non-linear first-order differential equation for $f$
\begin{equation}\label{e:C-H2}
f_1^2 +2 y_1 f_1 f_2 + 2 (y_1^2 -y_2 )f_2^2 = 1\,,
\end{equation}
where $f_i = \frac{\partial f}{\partial y_i}$ for  $i=1,2.$
The equation \re{e:C-H2} can be made to look simple by the change of variables
\begin{eqnarray}\label{e:vartrans}\left\{\begin{array}{lll}
y_1 &=& u_+ + u_- \\
y_2 &=& u^2_+ + u^2_-
\end{array}\right.\,.
\end{eqnarray}
Denoting the function derivatives by $f_\pm \equiv {\partial f\over \partial u_\pm}$, one has the remarkably simple first-order differential equation \cite{GR,PS,DS,XH}
\begin{equation}\label{duality}
f_+ f_-=1.
\end{equation}
We will refer to this equation enforcing duality symmetry as \emph{Courant-Hilbert equation}.

An other way to obtain the equation \re{duality} (as it was first obtained) \cite{GR} is to start from the usual one-potential Lagrangian formulation of non-linear electrodynamics, manifestly gauge and Lorentz invariant, with action $S[A_\m]=\int d^4x\, L(x,y).$
The function $L$ depends only on the two independent Lorentz scalars constructed from the curvature $F_{\m\n}$, namely $x = -\fr{1}{4}F_{\m\n}F^{\m\n}$ and $y =\fr{1}{64}(F_{\m\n}F^{*\m\n})^2.$
An important physical requirement is that we want to recover Maxwell theory in the weak field limit. 
For this, we require that $L(x,y)$ is \emph{analytic} in the neighborhood of $x=y=0$ and 
\begin{equation}\label{e:weakfield}
L(x,y)=x+O(x^2,y).
\end{equation}
To make link with equation \re{e:duality}, we make the change of variable
\begin{equation}\label{e:xy}
x = u_+ + u_- \,,\;\;
y = u_+ u_-.
\end{equation}
If we ask for (*) duality invariance of the equations of motion together with (**) the weak field limit \re{e:weakfield},
we obtain exactly the equation \re{duality} for the function $f(u_+,u_-):=L(x,y)$.
It is not surprising to find the same equation for the Hamiltonian density and the Lagrangian density because they are
related by a Legendre transformation, and Legendre transformation relates a model and its dual \cite{Zumino}.

This equation was also obtained for a chiral two-form in six dimensions, choosing as special the fifth direction
\cite{PS}, or the time direction \cite{XH} (in order to use the Dirac-Schwinger condition to enforce Lorentz invariance).
This comes from the previously mentioned fact that dimensional reduction of a chiral two-form from six to four dimensions
gives duality-symmetric electrodynamics, at the linear \cite{Verlinde} and non-linear \cite{Berman} level. We just mention that this equation was also obtained for a self-interacting massless scalar
field in four dimensions \cite{DS}.

\subsection{Self-couplings of the PST model} \label{ss:defPST}

Here we will seek for a deformed theory describing consistent self-couplings of  one Maxwell field (on-shell). Our model has to be Lorentz covariant generalizing the discussion from the previous part and the free PST model. Just as in the case of the free systems the covariantization involves an auxiliary field and an extra gauge invariance. As we will see below it is the deformation of this symmetry that will lead us to the Courant-Hilbert equation. 

It seems natural to require the interacting action to satisfy the same kind of symmetries as the free one. In other words we should expect besides the Lorentz invariance also a manifest duality-symmetry. If on top of that we ask that the interaction should depend only on the field strengths this reduces  the number of invariants to only two, namely
\begin{equation}
z_1 = \frac{1}{2}\tilde\cH^{\a}_p \tilde\cH^{\a p}\,, \; \;
z_2 = \frac{1}{4}\tilde\cH^{\a}_p \tilde\cH^{\b p}\tilde\cH^{\a}_q \tilde\cH^{\b q}\,.
\end{equation}
Thus, we propose as action for a self-interacting gauge vector 
\begin{eqnarray}
S^{I}&=& \int d^4x \, \left( \frac12 \cH^\a _m \tilde \cH^{\a m}  - f(z_1 ,z_2) \right) \,
\label{e:interpst}
\end{eqnarray}
where, similar to the non-covariant situation, we kept the ``kinetic" term and added an interaction term $f(z_1 ,z_2)$ depending only on  the invariants of the theory. Up to now $f$ is a general function but the connection with the free theory \re{e:pst} imposes its analyticity at the origin and its reduction to $f\ra z_1$ in the weak field limit. We are going to restrict the class of possible interations by demanding the field equations as well as the action to remain invariant under some modified transformations of type \re{e:ginv1}-\re{e:ginv3}. In fact we deform only the gauge symmetry \re{e:ginv2}\footnote{The other two gauge symmetries remain the same and play the same  role as in the free model.} to 

\begin{eqnarray}
\delta A_m^\a = \cL^{\b\a} (\cH_m^\b - J_m^\b )\fr{\f}{\sqrt{-u^2}}\,, &\,\delta a=\f \,,\label{e:defginv2}
\end{eqnarray}
where we denoted the deforming contribution as $J_m^\b = \frac{\d f}{\d \tilde \cH^{\b m}}\,.$ One can observe that in the free limit case the last transformation is nothing but \re{e:ginv2}.

Using the same approach as for the free case (i.e. gauge-fixing \re{e:ginv3}) the solution to the equations of motion determined by  $S^I$ read
\begin{eqnarray}
\cH^\a_{p} = J^\a_{p} = f_1 \tilde \cH^\a_{p} + f_2 (\tilde \cH^3)_p^\a \,,\label{e:defduality}
\end{eqnarray}
which should be understood as a generalization of the self-duality condition \re{e:duality}. Here $f_i = \frac{\partial f}{\partial z_i}$ for  $i=1,2$ and $(\tilde \cH^3)_p^\a = \tilde \cH^\b_{p} \tilde \cH^{\b q} \tilde \cH^\a_{q}$.

We want to find now what are the implications of the invariance of this equation under \re{e:defduality} on the function $f$. To see that one takes first its general variation, i.e.

\begin{eqnarray}
\d \cH^\a_{p} &=& f_1 \d  \tilde \cH^\a_{p} + f_2 \left(\d \tilde \cH^\b_{p} \tilde \cH^{\b q} \tilde \cH^\a_{q} + \tilde \cH^\b_{p} \d \tilde \cH^{\b q} \tilde \cH^\a_{q} + \tilde \cH^\b_{p} \tilde \cH^{\b q} \d \tilde \cH^\a_{q} \right) + \nn\\
&& + \tilde \cH^\a_{p} \left( f_{11} \d \cH^\b_{q} \tilde \cH^{\b q} + f_{12} \d \cH^\b_{q} (\tilde \cH^3)^{\b q} \right) + \nn\\
&& + (\tilde \cH^3)_p^\a \left( f_{12}\d \cH^\b_{q} \tilde \cH^{\b q} + f_{22} \d \cH^\b_{q} (\tilde \cH^3)^{\b q} \right) \,.\label{e:genvar}
\end{eqnarray} 
Instead of plugging in directly the transformation \re{e:defginv2} we first use the other symmetries of the system (Lorentz and SO(2) rotation invariance) to choose a basis in which the vector $u_p = \d_p^0$ (i.e. it is time-like) and the only non-vanishing components of the tensor $\tilde \cH^\a_{p}$ are 
\begin{equation}
\tilde \cH^1_{1} = \l_+ \,, \qq  \tilde \cH^2_{2} = \l_-\,.
\end{equation}
In this special choice the equation of motion reduces to only
\begin{eqnarray}
\g_{\pm} = \l_{\pm} (f_1 + f_2 \l_\pm^2) 
\end{eqnarray}
where $\g_+ =\cH^1_{1}$, $\g_- =\cH^2_{2}$ are the non-zero components of $\cH^\a_{p}$ in this basis.

This simplifies considerably the analysis of \re{e:genvar}. In fact there remain only two non-trivial possibilities for the choice of the $\a$, $p$ indices. The first one is $\a = 1$, $p=3$ (the situation $\a = 2$, $p=3$ is similar). It leads (up to some factors that cancel in \re{e:genvar} and after using the filed equations \re{e:defduality}) to the following variations $\d \cH^1_{3} = \l_-$, $\d \tilde\cH^1_{3} = \g_-$. Then \re{e:genvar} becomes 
\begin{displaymath}
\l_- = f_1 \g_-  + f_2 \g_- \l_+^2 \,,
\end{displaymath}
which upon using once more the equations of motion gives
\begin{equation}\label{e:C-H1}
(f_1 + f_2 \l^2_+)(f_1 + f_2 \l^2_-) = 1\,.
\end{equation}
One can reformulate it in terms of the two invariants $z_1 = \frac{1}{2} (\l^2_+ + \l^2_- )$ and $z_2 = \frac{1}{4}(\l^4_+ + \l^4_-)$ and performing the transformation \re{e:vartrans} one ends up with the Courant-Hilbert equation. Thus, we derived this equation by imposing the invariance of the equations of motion under the gauge transformation \re{e:defginv2}. 

The second non-trivial possibility resides in taking $\a =1$, $p=1$ or $\a =2$, $p=2$. A priori one should expect that the two conditions derived in this way  from \re{e:genvar} will lead to further constraints on the second-order derivatives of $f$. Nevertheless, one can show that, taking the sum and the difference of the two relations, one arrives to some consequences of \re{e:C-H2} (linear combinations of the derivatives of this equation). In other words the restriction \re{duality} is the only one that follows from the invariance of \re{e:defduality}.

We would like now to see explicitely that the action is as well invariant (up to total derivatives) under the same gauge transformations \re{e:defginv2}. Inserting \re{e:defginv2} in the general variation of the action
\begin{eqnarray}
\d S^I &=&  \int d^4x \, \left[ ( \cH^\a_{p} - J^\a_{p}) \d_{A,a} \tilde \cH^{\a p} +\frac{1}{2} \d_a \cH^\a_{p} \tilde \cH^{\a p} - \frac{1}{2} \cH^\a_{p} \d_a \tilde \cH^{\a p} \right]
\end{eqnarray} 
and, after canceling the contributions of  first-order in the derivatives of $f$ coming from the variation with respect to $A_m^\a$, respectively $a$, one deduces
\begin{eqnarray}\label{e:JJ}
\d S^I &=& \frac{1}{2} \int d^4x \, \e^{mnpq} \cL^{\b\a} \d v_m v_n \left( \tilde\cH^\a_{p} \tilde\cH^\b_{q} - J^\a_{p} J^\b_{q} \right)\,.
\end{eqnarray} 

The idea is to prove that the ``second-order" (in  the derivatives) term $JJ$ gives the same contribution as in the free theory thus, it cancels the other term. The simplest way to achieve that is to evaluate it in the special basis mentioned above. Indeed, in such a basis \re{e:JJ} becomes  
\begin{equation}
\d S^I = - 2 \int d^4x\, \d v_3 \,\l_+ \l_- \,[(f_1 + f_2 \l^2_+)(f_1 + f_2 \l^2_-)-1]\,.
\end{equation}
Therefore, upon applying the Courant-Hilbert equation \re{duality} one gets $\d S^I =0$, i.e. the invariance of the self-interacting action under the modified gauge transformations \re{e:defginv2}. The restriction \re{e:C-H1} is sufficient to guarantee the invariance of both, field equations and action.  

In conclusion, we have constructed a modified Lorentz covariant theory for the Maxwell field that possess also an electric-magnetic duality. The allowed self-interactions are restricted to a class of functions of two variables that must satisfy the Courant-Hilbert equation. The physical solutions of this equation are discussed in the next section. Moreover, it was shown that the deformed model has some modified gauge invariance, similar to the free case.

\subsection{Physical solutions to the Courant-Hilbert equation} \label{ss:solC-H}

As pointed in \cite{PS}, the general solution of \re{duality} has been given by Courant and Hilbert \cite{courant}\footnote{That is why we refer it as the Courant-Hilbert equation. Let us mention that the authors of
\cite{Hatsuda} gave an interesting alternative form of the general solutions.}.
The general solution is given implicitly in terms of an arbitrary function $z(t)$:
\begin{eqnarray}
f &=& {2u_+\over \dot z(t)} + z(t)\,,\nonumber \\
u_- &=& {u_+\over (\dot z(t))^2} + t,\label{e:general}
\end{eqnarray} 
where the dot means the derivative of the function with respect to its argument.
In principle, the second equation determines $t$ in terms of $u_+$ and $u_-$,
which can then be substituted into the first one to give $f$ in terms of $u_+$ and $u_-$.
Unfortunately, in practice this method to generate solution is not tractable for arbitrary $z(t)$.

Furthermore, we should not forget the requirement of analyticity at the origin on $L(x,y)$, as well as \re{e:weakfield}.
The theorem presented here says that these two requirements can be equivalently translated into a precise condition on the
generating function $z(t)$.
\begin{theorem}\label{t:analyticity}
Let $f(u_+,u_-)$ be a solution of $f_+f_-=1$.
The function $L(x,y)\equiv f\left(u_+(x,y),u_-(x,y)\right)$ 
\begin{itemize}
\item is analytic near $(x,y)=(0,0)$ and 
\item satisfies $L(x,y)=x+O(x^2,y)$,
\end{itemize}
if and only if the boundary condition $L(t,0) \equiv z(t)$ is such that the function $\Psi(t)\equiv
-t\dot{z}^2(t)$
\begin{itemize}
\item[(i)] is equal to its inverse: $\Psi\left(\Psi(t)\right)=t$, 
\item[(ii)] is distinct from the identity: $\Psi(t)\neq t$, 
\item[(iii)] is analytic near the origin $t=0$ and 
\item[(iv)] vanishes at the origin: $\Psi(0)=0$.
\end{itemize}
\end{theorem}
It has been shown by Perry and Schwarz that $(i)-(iv)$ were necessary, the proof that it is also sufficient is given in the
appendix \ref{l:analyticity}.

One of the main interest of the theorem \ref{t:analyticity} is to prove that \emph{there exists an infinite class of
physically relevant duality-symmetric theories} even if only one explicit example is known\footnote{Notice that the
Lagrangians given in \cite{Hatsuda} are not analytic at the origin in the variables $x$ and $y$. This point was raised to
our attention by \"{O}. Sar{\i}o\~{g}lu.}. 

To see that there exists an infinite class of them, we can go to the procedure given by Perry and Schwarz to generate
large class of solutions for
\begin{equation}
\Psi\left(\Psi(t)\right)=t.
\end{equation}
Let $F(s,t)$ be an analytic function near the origin such that $F(s,t)=s+t+O(|(s,t)|^2)$.
Then the implicit equation $F(s,t)=0$ defines a function $s=\Psi(t)=-t+O(t^2)$ analytic near the origin (application of
implicit function theorem \cite{Dieu}).
An interesting point is that if the function $F$ is symmetric, then the implicit function $\Psi(t)$ is equal to
its inverse because $F(s,t)=F(t,s)=0$
implies $t=\Psi(s)=\Psi\left(\Psi(t)\right)$. 

The simplest non-trivial example $F(s,t) = s + t + \alpha s t$ generates the Born-Infeld electrodynamics at the end of the whole procedure ($\a=0$ corresponds to Maxwell theory, which is not considered here as a distinct example of solution). 
Unfortunately, all this procedure becomes rapidly cumbersome and no other explicit example of duality-symmetric theory is known.
Anyway, our theorem shows that duality invariance together with analyticity is not enough to single out uniquely Born-Infeld theory, contrary to what could have been conjectured from the fact that only one explicit example is known. The theorem \ref{t:analyticity} ensures that we can generate implicitly an infinite class of analytic solutions at the origin.

\section{Conclusions}\label{s:concl}

In the present paper we completely classified all consistent local interactions of a system of $p$-form gauge fields (with $p$ odd)
described at the free level by a sum of non-covariant duality-symmetric actions in $2p+2$ dimensions. To handle this problem, we made use of the powerful tool of homological perturbation theory, by reformulating the problem in
terms of the local BRST cohomology. 

We found that no deformation of the gauge transformations is allowed, the only consistent interactions being of two types: either polynomials in the curvatures
and their spatial derivatives, or Chern-Simons like terms.
Of course the strength of a no-go theorem is directly proportional to the weakness of its hypothesis, in this case: 
(i) continuous and (ii) local deformations of a sum of duality-symmetric actions.
Hence, the absence of the analogue of a Yang-Mills type deformation suggests that non-Abelian duality requires to take into account unavoidable non-perturbative or non-local features (or even more exotic properties). For instance, the authors of \cite{Chan:1996} proposed a generalized duality symmetry for non-Abelian Yang-Mills fields (Their use of loop space formalism is presumably responsible for non-locality.).

Using the previous results, we pointed all the assumptions leading to the Courant-Hilbert equation enforcing duality symmetry for sourceless non-linear electrodynamics, either in Lagrangian or Hamiltonian formalism. Besides of the assumption of slowly varying field strength, we can say that the Courant-Hilbert relation determines entirely all possible Lorentz and duality-symmetric theories of electrodynamics. Afterwards, we deduced the same restriction on the possible self-couplings of the Maxwell field (only in 4 dimensions) in the Lorentz covariant PST formulation by requiring the invariance of the model under some deformed gauge symmetry. This has been done for simplicity in a flat Minkowski space but, due to the covariance, it can be directly generalized to a curved background, as it was discussed for a Born-Infeld interaction in 6 dimensions by the authors of \cite{M5}. It would be of course interesting to generalize these last results to higher dimensions. A step in that direction has been taken recently in \cite{fks} were the deformations of a supersymmetric Yang-Mills  theory in 10 dimensions have been investigated.

Perry and Schwarz derived a set of necessary conditions to generate physically relevant solutions to the Courant-Hilbert equation. Finally, we proved that this set is also sufficient, yielding to the conclusion that there exists an infinite class of physically relevant theories of duality-invariant non-linear electrodynamics.

\section*{Acknowledgments}

We thank M. Henneaux for suggesting the problem and for his support in this project. 
We are also grateful to G. Barnich, S. Benzel, P. Bieliavsky, N. Boulanger, P.-J. De Smet, B. Knaepen, \"{O}. Sar{\i}o\~{g}lu, W. Troost and A. Van Proeyen for useful discussions. We acknowledge the ``Centre \'Emile Borel" for their hospitality during the first part of this work.

This work was supported in part by the ``Actions de Recherche Concert{\'e}es" of the
``Direction de la Recherche Scientifique - Communaut{\'e} Fran{\c c}aise de Belgique",
by IISN - Belgium (convention 4.4505.86)
and by the European Commission TMR programme HPRN-CT-2000-00131, in which X.B. is associated to Leuven.

\noindent
\appendix

\section{Locality requirement for the deformations}\label{a:locality}

The following lemma is a particular case of standard techniques developed in \cite{Barnich:1995} concerning locality
requirement
(see Sec. 6 and 7 of the review \cite{Barnich:2000}).

\begin{lemma}\label{l:expansion}
Let $\a$ be a representative of an element of $H_0(s|d)$.
If its lowest component $\a_0$ in $antigh$
number $0$ is local and contains less than $k$ derivatives,
then we can assume that all the terms in the $antigh$ expansion of $\a$ are local. Furthermore, this series stops in {\it
antigh} number smaller than $k$.
\end{lemma}
\proof{Let us decompose the cocycle $\a$ according to the {\it antigh} number, i.e. $\a=\a_0+\a_1+\dots$ where $antigh(\a_i)=i$.
It is a standard result that the non-trivial cocycles $\a$ of $s$ modulo $d$ with {\it gh} number $g$
are in one-to-one correspondence with their lowest component $\a_g$ in {\it antigh} number, e.g. $\a_0$ here.

The operator counting the number of derivatives is defined by
\begin{equation}
N_\partial=\sum_l l\partial^{(l)}\Phi^A\frac{\partial}{\partial (\partial^{(l)}\Phi^A )}+\sum_l l
\partial^{(l)}\Phi^*_A\frac{\partial}{\partial (\partial^{(l)}\Phi^*_A )}.
\end{equation} 
We assume that $\a_0$ is a local function containing less than $k$ derivatives $N_\partial(\a_0)\leq k$.
Let us now show that we can assume $\a_l=0$ for $k<l$ without any loss of generality.
We introduce the degree $M$ counting the sum of the number of derivatives and
the number of antifields
\begin{equation}
M=\sum_l l\partial^{(l)}\Phi^A\frac{\partial}{\partial (\partial^{(l)}\Phi^A )}+\sum_l
(l+1)\partial^{(l)}\Phi^*_A\frac{\partial}{\partial(\partial^{(l)}\Phi^*_A)}.
\end{equation} 
One can verify that $M(\delta)=M(\gamma)=M(d)=1$.
The component of vanishing {\it antigh} number of the cocycle condition $s\a+d\b=0$ reads
\begin{equation}
\gamma\a_0 + \delta\a_1 + d\b_0=0 \,.
\end{equation}
Because all the operators have the same degree $M$, this last equation is homogeneous
in that degree and therefore we can always assume that we take a representative such that $M(\a_1)\leq k$.
Now, we can repeat this reasoning for the component of {\it antigh} number 1 of the Wess-Zumino
consistency condition to show that $M(\a_2)\leq k$. 
And so on until $M(\a_{k+1})\leq k$, which is not possible.

To control the locality in the expansion, we should introduce the degree $\bar{M}$ counting the sum of the number of
derivatives,
number of fields and twice the number of antifields:
\begin{equation}
\bar{M}=\sum_l (l+1)\partial^{(l)}\Phi^A\frac{\partial}{\partial (\partial^{(l)}\Phi^A )}+\sum_l
(l+2)\partial^{(l)}\Phi^*_A\frac{\partial}{\partial (\partial^{(l)}\Phi^*_A )}.
\end{equation} 
We can use the same argument as before, noticing that locality is equivalent
to the fact that the degree $\bar{M}$ is bounded.}

\section{Cohomology of $\gamma$}\label{a:gamma}

Let us first introduce some notations. Once this is done, it will be quite easy to formulate next lemma.

The algebra ${\mathcal{I}}$ is the algebra of local forms with coefficients that are polynomial
in the fields $F^{\a A}_{ij}$ together with all their spatial derivatives up to a finite order, and the fields 
$\bar{F}^{\a A}_{ij}\equiv \dot{F}^{\a A}_{ij} - \cL^{\a\b}\epsilon_{ijk}\partial _l F^{\b A\, kl}$
as well as all their partial derivatives up to a finite order.
The algebra $\Phi^*$ is the algebra of local functions in the antifields $\phi^*_A$. 
The algebra generated by $C^{\a A}$ and all their time derivatives up to a finite order is denoted by ${\mathcal{C}}$.

\begin{lemma}\label{l:gamma}
The cohomology of $\gamma$ is given by
\begin{equation}\label{e:gamma}
H(\gamma)={\mathcal I}\otimes{\Phi}^*\otimes{\mathcal C}.
\end{equation}
\end{lemma}
\proof{
By using the results of \cite{Barnich:1995,knaepen} about the cohomology of $\gamma$ for Abelian 1-forms,
we know that the generators of the invariant local functions in $H(\gamma)$ are the curvatures $F^{\a A}_{ij}$,
the antifields $\phi^*_A$ with all their (spacetime) derivatives up to a finite order,
the ghosts $C^{\a A}$ and all their time derivatives up to a finite order.
The lemma \ref{l:gamma} is obtained from an invertible change of variables from the set of generators $\{\partial^{}_{m_1 \ldots m_k}F^{\a A}_{ij}\}$ to the set $\{\partial^{}_{i_1 \ldots i_l}F^{\a A}_{ij},\,\partial^{}_{m_1 \ldots m_l}\bar{F}^{\a A}_{ij}\}$.
}

The generators of ${\mathcal I}$ are denoted collectively by $I$ and the generators of ${\mathcal I}^*\equiv{\mathcal I}\otimes
{\Phi}^*$ are labelled by $\chi$.
We can also select a basis $\{\omega^I\}$ of ${\mathcal C}$ (the index $I$ goes from $1$ to $dim({\mathcal C})$).
If $\alpha$ is a local form then theorem \ref{l:gamma} tells that
\begin{equation}
\gamma \alpha =0 \quad\Leftrightarrow \quad\alpha =\sum_I P_I(\chi)\omega^I+ \gamma\beta.
\end{equation}
Furthermore, the following property is used several times in this paper
\begin{equation}
\sum_I P_I(\chi)\omega^I=\gamma\beta \quad\Rightarrow \quad P_I(\chi)=0.\label{glll}
\end{equation}
Indeed, no non-vanishing element of $H(\gamma)$ can be $\gamma$-exact thus
$\sum_I P_I(\chi)\omega^I$ has to vanish, and then the property immediately follows because $\omega^I$ is a basis.


\section{Algebraic and invariant Poincar\'e lemmas}\label{a:Poincar}

In order to prove and use the different Poincar\'e lemmas, we will assume in the sequel to work with a the spacetime manifold $M$ having a trivial topology. Otherwise, the statements would be valid only locally. This requirement is consistent with the topological assumptions necessary to obtain the (on-shell) equivalence with Maxwell theory.

The following lemma is called the algebraic Poincar\'e lemma (for $q<4$).
\begin{lemma}\label{l:Poincalg}
The cohomology of $d$ in the algebra of local forms of degree $q<4$ is given by
\begin{eqnarray}
H^0(d) &\simeq& \Rrbar\,, \nonumber\\
H^q(d) &=& 0\,, \quad\rm{ for } \,\, q\not = 0.\nonumber
\end{eqnarray}
\end{lemma}
A proof of this lemma can be found in \cite{Vinogradov}.

The next lemma still concerns the cohomology of $d$ but for the local forms belonging to the space ${\mathcal I}^*$.
This lemma is called the invariant Poincar\'e lemma because the restricted space is invariant (and non-trivial) under the
action of $\gamma$. 
\begin{lemma}\label{l:invP}
Let $P^q(\chi)$ be a local form of degree $q<4$, then
\begin{equation}
d\, P^q(\chi)=0\Leftrightarrow P^q(\chi)=\lambda + dx^0\tilde{P}^{q-1}\left((\partial_0)^aF^{\a A}\right)+dQ^{q-1}(\chi),
\end{equation}
where $\lambda$ is a constant and $\tilde{P}^{q-1}\left((\partial_0)^aF^{\a A}\right)$ is a polynomial in the curvature forms $F^{\a A}=\frac12 F^{\a A}_{ij}\,dx^i
dx^j$ and all its time derivatives up to a finite order. 
\end{lemma}

For $q>0$, we define ${\mathcal F}^q$ as the space of local $q$-forms that are the product of $dx^0$ with any polynomial
$\tilde{P}^{q-1}\left((\partial_0)^aF^{\a A}\right)$ in the $F^{\a A}$ and all its time derivatives up to a finite order,
defined modulo total time derivatives: $(dx^0 \partial_0)\tilde{Q}^{q-1}\left((\partial_0)^aF^{\a A}\right)$.
We can refine the previous lemma.
For $q<4$, the invariant cohomological groups of $d$ in antighost number $k$, $H_k^{q,inv}(d) \equiv H_k^q(d,{\mathcal I}^*)$, are given by
\begin{eqnarray}
H_k^{q,inv}(d)&=&0 \,, \quad\rm{for}\,\, k\not= 0\,,\nonumber\\
H_0^{q,inv}(d)&=&{\mathcal F}^q \,, \quad\rm{for}\,\, 0<q<4\,,\nonumber\\
H_0^{0,inv}(d)&=&\Rrbar \,.\\
\end{eqnarray}
\proof{The proof of the case $antigh$ number $k\neq 0$ is identical to the standard one, which can be found for instance in
\cite{Barnich:1995,Barnich:2000,knaepen}.
Thus, we just have to examine the cocycle condition $dP^q(I)=0$ for $q<4$. Any $q$-form can be decomposed into a sum of a
purely spatial $q$-form and one involving $dx^0$, i.e. $P^q(I)=\tilde{P}^q(I)+dx^0\tilde{Q}^{q-1}(I)$. Using an appropriate basis of ${\mathcal C}$ (see \cite{Bekaert:1999}), it is possible to show that the cocycle condition implies $\tilde{P}^q(I)=\lambda+\tilde{d}\tilde{R}^{q-1}(I)$, 
hence $P^q(I)=\lambda+dx^0\tilde{Q}^{q-1'}(I)+d\tilde{R}^{q-1}(I)$ with
$\tilde{Q}^{q-1'}(I)=\tilde{Q}^{q-1}(I)-\partial_0\tilde{R}^{q-1}(I)$.
The cocycle condition is then equivalent to $\tilde{d}\tilde{Q}^{q-1'}(I)=0$.

Before analyzing this last equation we recall the \textit{spatial} invariant Poincar\'e lemma, that is
\begin{eqnarray}
&&\tilde{d}\tilde{R}^q(\chi)=0\Leftrightarrow
\tilde{R}^q(\chi)=\tilde{R}^q\left((\partial_0)^aF^{\a A}\right)+\tilde{d}\tilde{S}^{q-1}(\chi)\,,\\
&& \tilde{R}^q\left((\partial_0)^aF^{\a A}\right) = \tilde{d}\tilde{T}^{q-1}(I) \Rightarrow
\tilde{R}^q\left((\partial_0)^aF^{\a A}\right) = 0
\end{eqnarray}
for $q<3$. To prove it, we first look at the generators of the algebra ${\mathcal I}$
\begin{equation}
\{\partial^{}_{i_1 \ldots i_k}(\partial_0)^aF^{A}_{ij},\,x^m,\,dx^m\}\,,
\end{equation}
where $k$ and $a$ are positive integers.
Considering $a$ and $A$ as only one label and forgetting $x^0$ and $dx^0$, this set is exactly the same as the corresponding set of generators of the algebra $H(\gamma)$ in vanishing $puregh$ number for a system of spatial one-forms $A_i^M\equiv (\partial_0)^a A_i^{\a A}$ in 3 dimensions. Furthermore, $dx^0$ is not present here, and $x^0$ plays the same role as a constant.
In conclusion, we can simply use the results obtained in \cite{knaepen} for the cohomology group $H(\gamma|\tilde{d})$ of a system of $1$-forms in any dimension.

Applying the spatial invariant Poincar\'e lemma to $\tilde{d}\tilde{Q}^{q-1'}(I)=0$, we find that
$P^q(I)=\lambda+dx^0\tilde{Q}^{q-1}\left((\partial_0)^aF^{\a A}\right)+dT^{q-1}(I).$ Now, let us assume that $P^q(I)$ is trivial ($q$ is still $<4$) 
\begin{equation}
\lambda=0\,,\quad dx^0\tilde{Q}^{q-1}\left((\partial_0)^aF^{\a A}\right)=dT^{q-1}(I).
\end{equation}
With the decomposition $T^{q-1}=\tilde{T}^{q-1}+dx^0\tilde{U}^{q-2}$, the triviality condition implies
$\tilde{d}\tilde{T}^{q-1}(I)=0$. From spatial invariant Poincar\'e lemma we deduce, in the same way as before, that we can assume
$T^{q-1}=\tilde{T}^{q-1}\left((\partial_0)^aF^{\a A}\right)+dx^0\tilde{U}^{q-2'}(I)+d\tilde{V}^{q-2}(I)$.
Finally, we get 
\begin{equation}
\tilde{Q}^{q-1}\left((\partial_0)^aF^{\a A}\right)+\partial_0\tilde{T}^{q-1}\left((\partial_0)^aF^{\a A}\right)=\tilde{d}\tilde{U}^{q-2'}(I).
\end{equation}
The spatial invariant Poincar\'e lemma gives
\begin{equation}
\tilde{Q}^{q-1}\left((\partial_0)^aF^{\a A}\right)+\partial_0\tilde{T}^{q-1}\left((\partial_0)^aF^{\a A}\right)=0\,,
\end{equation}
which means that the only possibility for the curvatures terms to be trivial is that they are the time derivatives of something, as
stated before.}

\section{Cohomology of $\gamma$ modulo $d$}\label{a:gammamodd}

Let be $\a^p$ a local $p$-form satisfying
\begin{equation}
\gamma \a^p + d\b^{p-1}=0.
\label{e:gammad}
\end{equation}
From (\ref{e:gammad}), applying the differential $\gamma$ and using the algebraic Poincar\'e lemma \ref{l:Poincalg}, we derive the descent equations
\begin{eqnarray}
\gamma \b^{p-1} &+&d\b^{p-2}=0 \,,\\
&\vdots& \,,\nonumber\\
\gamma \b^{q+1}&+&d\b^q=0 \,,\label{e:before}\\
\gamma \b^{q}&=&0 \,.\label{e:last}
\end{eqnarray}
The descent possesses a bottom of the form (\ref{e:last}) as the form degree is positive.

Let us suppose $q<p$. The equation (\ref{e:last}) implies that $\b^{q}=\b_I^{q}(\chi)\omega^I+\gamma\rho^q$ (see App. \ref{a:gamma}).
Inserted in (\ref{e:before}), this yields $d\b_I^{q}(\chi)=0$. From the invariant Poincar\'e lemma \ref{l:invP}, we deduce also $\b^{q}=\lambda+dx^0P_I^{q-1}\left((\partial_0)^aF^{\a A}\right)\omega^I+\gamma\rho^q+d\sigma^{q-1}$.
The trivial part in $H^q(\gamma|d)$ can be removed by a $d$-trivial redefinition of $\b^{q+1}$ that does not affect the
descent equation before equation (\ref{e:before}).
If $\b^q$ is trivial, then the bottom is really one step higher.

All this enables us to transpose the results valid for spatial $1$-forms in $3$ dimensions (that can be found in
\cite{Barnich:1995,knaepen}) to our problem. We thus obtain the following lemmas.

\begin{lemma}\label{l:gammamodd0}
The elements of $H^4_{0,\,0}(\gamma|d)$ are of two types:
\begin{itemize}
\item[A.]{those which are strictly invariant under $\gamma$, they are polynomials in curvatures and their partial
derivatives, i.e.
of the form 
\begin{equation}f(\partial_{m_1\dots m_k}F^{\a A}_{ij})\,d^4x\,;\end{equation}} 
\item[B.]{those which are invariant only up to a total derivative,
they are linear combinations of Chern-Simons like terms, i.e. 
\begin{equation}\lambda_{\a\b\,ab\,AB}\,(\partial_0)^aA_i^{\a A}(\partial_0)^bB^{\b B\, i} \,d^4x\,,\end{equation}
where $\lambda_{ab\,\a\b\,AB}$ are some constants such that 
$\lambda_{ba\,\b\a\,BA}=\lambda_{ab\,\a\b\,AB}$, in order for the cocycle not to be
a trivial boundary term.}
\end{itemize} 
\end{lemma}

\begin{lemma}\label{l:gammamodd'}
Let $\a$ be a local form of non-vanishing $antigh$ number that fulfills
$\gamma \a + d\b=0$. There exist a local form $\b'$ so that $\a':=\a+d\b'$ satisfies $\gamma \a'=0$.
\end{lemma}

\begin{lemma}\label{l:gammamodd''}
There is no element of the group $H^3(\gamma|d)$ in $puregh$ number 1 and $antigh$ number 0, that descends non-trivially.
\end{lemma}

\section{Invariant cohomology of $\delta$ modulo ${\tilde d}$}

The derivation of the invariant cohomology of $\delta$ modulo $\tilde{d}$ is essentially based on usual perturbative arguments. 
Let us introduce the operator $N$, counting the sum of the number of spatial derivatives and the number of antifields (with
equal weight). This operator $N$ is the derivation defined by
\begin{eqnarray}
&& N(\Phi^{*})=\Phi^{*}\,,\quad N(\Phi)=0\,,\\
&& N(\partial_k)=\partial_k\,,\quad N(\partial_0)=0\,,\\
&& N(x^m)=N(dx^m)=0\,,
\end{eqnarray}
where $\Phi$ a,d $\Phi^*$ denote respectively the fields and the antifields.
The degree $N$ of $\tilde{d}$ is obviously 1. 
The Koszul-Tate differential decomposes as $\delta=\delta_0+\delta_1$.
The differential $\delta_0$ acts non-trivially only on the antifields $A^{\a A*}_i$.
The differential $\delta_1$ acts on the generators in the same way as the Koszul-Tate differential
of a system of spatial Abelian $1$-forms in 3 dimensions,
for which we know the results of \cite{Barnich:1995,knaepen}.
Among other things they tell us that

\begin{lemma}\label{l:delta1}
The cohomology groups $H_k^{3,inv}(\delta_1|\tilde{d})$ are trivial for $k>2$,
and the only non-trivial elements of $H_2^{3,inv}(\delta_1|\tilde{d})$
are linear combinations of forms \newline $(\partial_0)^a C^{\a A*} d^3x$.
\end{lemma} 

The results of the previous lemma can be extended 
to the corresponding cohomology groups $H^{3,inv}_k(\delta|\tilde{d})$
because

\begin{lemma}\label{l:delta0+1}
$H_k^{3,inv}(\delta|\tilde{d})\cong H_k^{3,inv}(\delta_1|\tilde{d})$ for $k>1$.
\end{lemma}
\proof{Let $\a(\chi)$ be an invariant cocycle of $\delta$ modulo $\tilde{d}$ of {\it antigh} number $k>1$ and form degree
$3$, i.e. $\delta\a(\chi)+\tilde{d}\b(\chi)=0$.
Its expansion in degree $N$ is $\a=\a_1(\chi)+\dots+\a_n(\chi)$, $N(\a_i)=i\a_i$.
The cocycle condition in $N$-degree equal to $n+1$ is $\delta_1\a_n(\chi)+\tilde{d}\b_n(\chi)=0$. From lemma \ref{l:delta1} we deduce that $\a_n(\chi)=a_n(\chi)+\delta_1b_{n-1}(\chi)+\tilde{d}c_{n-1}(\chi)$,
where $a_n(\chi)$ is a non-trivial cocycle of $H_k^{3,inv}(\delta_1|\tilde{d})$ (only present for $k=2$ and $n=1$)
which does not contain the antifields $A^{\a A*}_i$. Therefore, $a_n(\chi)$ is also a cocycle of
$H_k^{3,inv}(\delta|\tilde{d})$. 
Let us define $\a^{(1)}:=\a-a_n-\delta b_{n-1}+\tilde{d}c_{n-1}$.
It is an invariant cocycle of $\delta$ modulo $\tilde{d}$ and its decomposition stops in $N$-degree $n-1$ ($\a^{(1)}_i=0$
for $i\geq n$). We can apply the same reasoning until we arrive at $\a^{(n)}=0$, which happens after at most $n-1$
supplementary steps
(\textit{remark}: 
This works because $\delta_1 b_0=\delta b_0$, 
coming from $\delta_0\Phi^A=0$).
Eventually, we have $\a=a(\chi)+\delta b(\chi) + \tilde{d} c(\chi)$
with $a(\chi)$ a non-trivial cocycle of $H_k^{3,inv}(\delta_1|\tilde{d})$.
It is also non-trivial in $H_k^{3,inv}(\delta|\tilde{d})$ because it does not involve spatial derivative 
and $\delta$ as well as $\tilde{d}$ increase the number of spatial derivatives.
Thus, the basis of non-trivial elements of $H_k^{3,inv}(\delta_1|\tilde{d})$ given in lemma \ref{l:delta0+1} provides
a basis of of non-trivial elements of $H_k^{3,inv}(\delta|\tilde{d})$.}

\begin{lemma}\label{l:lastl}
$H_1^{3,inv}(\delta|\tilde{d})$ is trivial.
\end{lemma}
\proof{Let us introduce the degree $\bar{N}$ counting the polynomiality in the fields 
$\bar{F}^{\a A}_{ij}$ and their partial derivatives
\begin{equation}
\bar{N}=\sum_l \partial^{(l)}\bar{F}^{\a A}_{ij}\frac{\partial}{\partial(\partial^{(l)}\bar{F}^{\a A}_{ij})}.
\end{equation}
As a consequence of $\delta A^{\a A*}_i=\fr12\cL^{\a\b}\epsilon_{ijk}\bar{F}^{\b A\,jk}$, the differential $\delta$ acting on $A^{\a A*}_i$ increases the degree $\bar{N}$ by one.

Then take $\a$ to be a representative of an element of $H_1^{3,inv}(\delta|\tilde{d})$. It is linear in the generators $\partial_{m_1}\dots\partial_{m_k}A^{\a A*}_i$ (with coefficients in $\mathcal{I}$). The cocycle condition $\delta\a(\chi)+\tilde{d}\b(I)=0$ can be decomposed according to the degree $\bar{N}$
\begin{eqnarray}\label{e:set}
&&\delta\a_i(\chi)+\tilde{d}\b_{i+1}(I)=0\,,\quad (i=0,\dots,k)\\
&&\bar{N}(\a_i)=i\a_i\,,\quad \bar{N}(\b_i)=i\b_i\,,
\end{eqnarray}
For $i\geq 0$ it is true that $\b_{i+1}(I)\approx 0$, as $\bar{F}^{\a A}_{ij}\approx 0$. It results 
\begin{equation}\label{e:insert}
\b_{i+1}(I)=\delta \m_i(\chi)\quad (i=0,\dots,k).
\end{equation}

We define $\mu(\chi)$ and $\a'(\chi)$ as $\m:=\m_0+\dots+\m_k$ and
$\a'(\chi):=\a(\chi)-\tilde{d}\m(\chi)$.
By inserting the equations (\ref{e:insert}) in (\ref{e:set})
we find that $\a'$ is a cocycle of the differential $\delta$ ($\delta\a'(\chi)=0$).
But a well-known theorem (see e.g. \cite{Barnich:1995,Barnich:2000}) states that the cohomology
of the Koszul-Tate operator $H_j(\delta)$ is trivial in positive {\it antigh} number $j>0$ for any gauge theory.
Hence, $\a'(\chi)=\delta \nu$. It is straightforward to show that we can assume $\nu$ invariant and
in conclusion, $\a=\delta\nu(\chi)+\tilde{d}\mu(\chi)$. This completes the proof.}

\section{Analyticity of solutions to Courant-Hilbert equation}\label{l:analyticity}

We present here a number of lemmas necessary in the discussion of the Courant-Hilbert equation (Subsec. \ref{ss:solC-H}). 
The first lemma of this appendix allows us to reformulate the analyticity of $L(x,y)$ as the analyticity of the function $f(u_+,u_-)$ together with its
symmetry property. The function $f(u_+,u_-)$ is only an intermediate tool necessary to achieve the proof of theorem \ref{t:analyticity}. The second lemma provides a necessary and sufficient requirement on the function $z(t)$ to generate symmetric functions while the third one relates the analyticity of $f(u_+,u_-)$ at the origin with the analyticity of $z(t)$ at the origin, giving also the behaviors of $f$ and $z$.
All together, these lemmas lead us to the theorem \ref{t:analyticity}.

Let $u_+$ and $u_-$ be the two roots of the second order polynomial in $u$
\begin{equation}
u^2-x\,u+y=(u-u_+)(u-u_-)=0\,,
\end{equation}
which is equivalent to \re{e:xy}.
The Newton's theorem on symmetric polynomials \cite{Newton} states, in particular, that any symmetric polynomial
$P(u_+,u_-)=P(u_-,u_+)$ in the roots $u_+$ and $u_-$ can be re-expressed as a polynomial
$Q(x,y):=P\left(u_+(x,y),u_-(x,y)\right)$ in the coefficients $x$ and $y$. 
The following lemma provides a generalization of this last property for functions of two variables, analytic at the
origin\footnote{This point was stressed by Marc Henneaux.}.
\begin{lemma}\label{l:analytic}
A function $L(x,y)$ is analytic at the origin $(x,y)=(0,0)$ if and only if the symmetric function 
\begin{equation}
f(u_+,u_-):=L\left(x(u_+,u_-),y(u_+,u_-)\right)
\end{equation} is analytic at the origin $(u_+,u_-)=(0,0)$.
\end{lemma}
Due to Newton's theorem, this lemma is obvious for formal power series but convergence matters are rather intricate to handle from that point of view.
\proof{The analyticity of $f(u_+,u_-)$ at the origin is of course necessary since the composition of two analytic functions in a neighborhood is also an analytic
function (in the corresponding neighborhood), and the functions $x(u_+,u_-)$ and $y(u_+,u_-)$ are analytic at the origin. 

To prove that the analyticity of $f(u_+,u_-)$ at the origin is sufficient to ensure that $L(x,y)$ is also analytic at
that point,
we make the following change of variables\footnote{We acknowledge Steven Benzel for useful disscusions at this
point.}: $x=u_++u_-$, $z=u_+-u_-$. This is a diffemorphism everywhere,
thus the function 
\begin{equation}
h(x,z):=f\left(u_+(x,z),u_-(x,z)\right)
\end{equation}
is analytic at the origin $(x,z)=(0,0)$. But the symmetry of $f(u_+,u_-)$ implies that
$h(x,z)$ is even in $z$, i.e. $h(x,z)=h(x,-z)$. Therefore, $h$ is only a function of $z^2$: $h(x,z)=h(x,z^2)$.
But $z^2=x^2 - 4 y$ is analytic function of $x$ and $y$, hence
\begin{equation}
L(x,y):=h\left(x(x,y),z^2(x,y)\right)
\end{equation}
is analytic at the origin $(x,y)=(0,0)$.
}
\begin{lemma}
A necessary and sufficient condition for the symmetry of the function $f(u_+,u_-)$ defined implicitly by (\ref{e:general})
is that the function \begin{equation}\Psi(t)\equiv -t\dot{z}^2(t)\end{equation} is equal to its inverse:
$\Psi\left(\Psi(t)\right)=t$.
\end{lemma}
\proof{It has been proved in \cite{PS} that it is necessary. One can argue as follows that the requirement on $z$ to satisfy $\Psi\left(\Psi(t)\right)=t$ suffices to have $f(u_+,u_-)$ symmetric.
We begin by noticing that this last requirement implies
\begin{equation}\label{e:z}
\dot{z}\left(-t\dot{z}^2(t)\right) \dot{z}(t)=\pm 1 
\end{equation}
Taking \re{e:z} at $t=0$ will select the positive sign.
Consider $f(u_+,u_-)$ the function defined by (\ref{e:general}).
We have to show that this function is symmetric.
From
\begin{equation}
u_{+}=\dot{z}^{2}(t)\left( u_{-}-t\right)  \label{S17}
\end{equation}
together with (\ref{e:z}), we deduce that 
\begin{equation}
u_{+}
=\frac{u_{-}}{\dot{z}^{2}\left( s\right) }+s  \,,\label{S14}
\end{equation}
with $s:=\Psi(t)$. With the help of (\ref{e:z}) and (\ref{S17}) we also get
\begin{equation}
f
=\frac{2u_{-}}{\dot{z}\left( -t\dot{z}^{2}(t)\right) }-2t\dot{z}(t)+z(t) \,. \label{S13}
\end{equation}
By taking the derivative, it can be checked that
\begin{equation}
z(t)-2t\dot{z}(t)=z\left( -t\dot{z}^{2}(t)\right) +K  \label{S11}
\end{equation}
where $K$ is a constant. By taking (\ref{S11}) at $t=0$, we find
that $K$ vanishes. Combining (\ref{S14}), (\ref{S13}) and (\ref{S11}) together we infer 
the symmetry of $f$, i.e.
\begin{equation}
f\left( u_{+},u_{-}\right) =f\left( u_{-},u_{+}\right) \equiv \left\{ 
\begin{array}{c}
f=\frac{2u_{-}}{\dot{z}\left( s\right) }+z\left( s\right) \\ 
u_{+}=\frac{u_{-}}{\left( \dot{z}\left( s\right) \right) ^{2}}+s
\end{array}
\right. \,.
\end{equation}}
\begin{lemma}\label{l:z}
Let $f(u_+,u_-)$ be a function defined implicitly by (\ref{e:general}).
It satisfies the following conditions
\begin{itemize}
\item[(i)] analyticity near $(u_+,u_-)=(0,0)$,
\item[(ii)] $f(u_+,u_-)=u_+ + u_- +O(|u_{\pm}|^2)$, 
\end{itemize}
if and only if the generating function $z(t)$ is analytic near $0$ and $z(t)=t+O(t^2)$.
\end{lemma}
\proof{The necessary condition is trivial and we focus immediately on the proof of the sufficiency.
Let us now define the function
\begin{equation}
F(u_+,u_-,t) \equiv u_{-}-\frac{u_{+}}{\left( \dot{z}(t) \right) ^{2}}-t \,.
\end{equation}
The function $\frac{1}{\left( \dot{z}(t) \right) ^{2}}$ is analytic near $t=0$ as $z(t)$ is also (near $t=0$) and
$\dot{z}(0)=1$. Hence, $F(u_+,u_-,t)$ is analytic near $(0,0,0)$. Furthermore,
\begin{equation}
\frac{\partial F}{\partial t}(u_+,u_-,t)=\frac{2u_{+} \ddot{z}(t)}{\left( \dot{z}(t) \right) ^{3}}-1 \rightarrow
\frac{\partial F}{\partial t}(0,0,0)=-1\neq 0\,.
\end{equation}
From standard theorems on implicit function \cite{Dieu}, we find that the function $g(u_+,u_-)$ defined by
$F\left(u_+,u_-,g(u_+,u_-)\right) =0$ is analytic near $(u_+,u_-)=(0,0)$.
The behavior of $g$ near the origin is $g(u_+,u_-)= u_- - u_+ +O(|u_{\pm}|^2)$.
Putting everything together one derives the analyticity of
\begin{equation}
f(u_+,u_-)\equiv\frac{2u_{+}}{\dot{z}\left( g(u_+,u_-)\right) }+z\left( g(u_+,u_-)\right)
\end{equation}
near $(u_+,u_-)=(0,0)$ and $f(u_+,u_-)=u_+ + u_- +O(|u_{\pm}|^2)$.
}

To finish with the proof of theorem \ref{t:analyticity}, we show that the four conditions: $\Psi\left(\Psi(t)\right)=t$, $\Psi(t)\neq t$, $\Psi(0)=0$ 
and $z(0)=0$ imply the good behavior of $z(t)$, i.e. 
$$z(t)=t+O(t^2),$$ as in lemma \ref{l:z}. Indeed, if we take the derivative of the first
condition, we get $\dot{\Psi}\left(\Psi(t)\right)\dot{\Psi}(t)=1$.
This implies $\left(\dot{\Psi}(0)\right)^2=1$ (using the third condition), hence $\dot{\Psi}(0)=\pm 1$. The
positive sign corresponds to the identity\footnote{The uniqueness of this solution can be seen from a simple
symmetry argument with respect to the diagonal in the $(t,\Psi)$-plane.} $\Psi(t)=t$ that we discard due to the second condition.
The negative sign together with $\Psi(0)=0$ gives: $\Psi(t)=-t\Phi(t)$ with $\Phi(t)=1+O(t)$.
Finally, from $\dot{z}^2(t)=\Phi(t)=1+O(t)$ and $z(0)=0$, we arrive at the expected conclusion: $z(t)=t+O(t^2)$. 

In conclusion, glueing together lemmas \ref{l:analytic}-\ref{l:z} with the last remark, we can see that the analyticity conditions on Lagrangian density $L(x,y)$ that generates duality-invariant equations of motion are equivalent to analyticity conditions on the function $z(t)=L(t,0)$. This was precisely the subject of theorem \ref{t:analyticity}.


\newpage

\begin{thebibliography}{99}


\bibitem{z:71} D. Zwanziger, 
Phys. Rev. {\bf D3} (1971) 880.

\bibitem{dt} S. Deser and C. Teitelboim, 
Phys. Rev. {\bf D13} (1976) 1592.

\bibitem{ss} J.H. Schwarz and A. Sen, 
Nucl. Phys. {\bf B 411} (1994) 35.

\bibitem{McClain:1990}
B.~McClain, F.~Yu and Y.~S.~Wu,
Nucl.\ Phys.\ B {\bf 343} (1990) 689;
N.~Berkovits,
Phys.\ Lett.\ B {\bf 395} (1997) 28,
[hep-th/9610134].

\bibitem{pst}
P.~Pasti, D.~ Sorokin and M.~Tonin,
Phys. Rev. {\bf D52} (1995) 4277, [hep-th/9506109];
\\
P.~Pasti, D.~ Sorokin and M.~Tonin,
 Phys. Lett. {\bf B352} (1995) 59, [hep-th/9503182].


\bibitem{Gu}
C.~Gu and C.~Yang,
Sci.\ Sin.\  {\bf 18} (1975) 483.

\bibitem{Barnich:1993}
G.~Barnich and M.~Henneaux,
Phys.\ Lett.\  {\bf B311} (1993) 123,
[hep-th/9304057];\\
M.~Henneaux,
\textit{Consistent interactions between gauge fields: The cohomological  approach},
Talk given at {\it Secondary Calculus and Cohomological Physics}, Moscow, Russia, 1997 [hep-th/9712226].

\bibitem{Bekaert:1999}
X.~Bekaert, M.~Henneaux and A.~Sevrin,
Phys.\ Lett.\  {\bf B468} (1999) 228,
[hep-th/9909094];\\
X.~Bekaert, M.~Henneaux and A.~Sevrin,
Nucl.\ Phys.\ Proc.\ Suppl.\  {\bf 88} (2000) 27
[hep-th/9912077];\\
X.~Bekaert, M.~Henneaux and A.~Sevrin, [hep-th/0004049].

\bibitem{Verlinde}
E.~Verlinde,
Nucl.\ Phys.\  {\bf B455} (1995) 211,
[hep-th/9506011];
D.~Berman,
Phys.\ Lett.\  {\bf B403} (1997) 250,
[hep-th/9612191].

\bibitem{BI}
M. Born and L. Infeld, Proc. Roy. Soc. (Lond) {\bf A144} (1934) 425.

\bibitem{Schr}
E. Schr\"odinger, Proc. Roy. Soc. (Lond) {\bf A150} (1935) 465.

\bibitem{Zumino}
M.~K.~Gaillard and B.~Zumino,
Nucl.\ Phys.\  {\bf B193} (1981) 221.

\bibitem{GR}
G.~W.~Gibbons and D.~A.~Rasheed,
Nucl.\ Phys.\  {\bf B454} (1995) 185,
[hep-th/9506035].

\bibitem{PS} 
M. Perry and J. H. Schwarz,
Nucl. Phys. {\bf B489} (1997) 47, [hep-th/9611065];\\
J. H. Schwarz, Phys. Lett. {\bf B395} (1997) 191, [hep-th/9701008].


\bibitem{DS}
S.~Deser and O.~Sarioglu,
Phys.\ Lett.\  {\bf B423} (1998) 369,
[hep-th/9712067].

\bibitem{XH}
X.~Bekaert and M.~Henneaux,
Int.\ J.\ Theor.\ Phys.\  {\bf 38} (1999) 1161,
[hep-th/9806062].

\bibitem{Sorin}
X. Bekaert and S. Cucu, 
JHEP{\bf 0101} (2001) 015,
[hep-th/0010266].

\bibitem{Piguet}
O.~M.~Del Cima, O.~Piguet and M.~S.~Sarandy,
[hep-th/0012067].

\bibitem{uniqueness}
F.~A.~Berends, G.~J.~Burgers and H.~van Dam,
Nucl.\ Phys.\  {\bf B260} (1985) 295;\\
R.~M.~Wald,
Phys.\ Rev.\  {\bf D33} (1986) 3613;\\
G.~Barnich, M.~Henneaux and R.~Tatar,
Int.\ J.\ Mod.\ Phys.\  {\bf D3} (1994) 139
[hep-th/9307155].

\bibitem{BernardK}
M.~Henneaux and B.~Knaepen,
Phys.\ Rev.\ D {\bf 56} (1997) 6076,
[hep-th/9706119];\\
M.~Henneaux and B.~Knaepen,
Int.\ J.\ Mod.\ Phys.\ A {\bf 15} (2000) 3535,
[hep-th/9912052].

\bibitem{bv} I.A. Batalin and G. A. Vilkovisky, 
\textit{Quantization of gauge theories with linearly dependent generators,} 
Phys. Rev. {\bf D28} (1983), 2567.

\bibitem{Barnich:1995}
G.~Barnich, F.~Brandt and M.~Henneaux,
Commun.\ Math.\ Phys.\  {\bf 174} (1995) 57,
[hep-th/9405109];
\\
G.~Barnich, F.~Brandt and M.~Henneaux,
Commun.\ Math.\ Phys.\  {\bf 174} (1995) 93,
[hep-th/9405194].

\bibitem{Barnich:2000}
G.~Barnich, F.~Brandt and M.~Henneaux,
Phys.\ Rept.\ {\bf 338} (2000) 439,
[hep-th/0002245].

\bibitem{knaepen}
M. Henneaux, B. Knaepen and C. Schomblond, Commun. Math. Phys. {\bf 186} (1997) 137 [hep-th/9606181];\\
M. Henneaux and B. Knaepen, Nucl. Phys. {\bf B548} (1999) 491, [hep-th/9812140].

\bibitem{Berman}
D.~Berman,
Phys.\ Lett.\  {\bf B409} (1997) 153, [hep-th/9706208];\\
A.~Nurmagambetov,
Phys.\ Lett.\ {\bf B436} (1998) 289,
[hep-th/9804157].

\bibitem{courant}  R. Courant and D. Hilbert, ``Methods of Mathematical
Physics,'' Vol. II (Interscience, 1962), p. 91.

\bibitem{Hatsuda}
M.~Hatsuda, K.~Kamimura and S.~Sekiya,
Nucl.\ Phys.\  {\bf B561} (1999) 341,
[hep-th/9906103].

\bibitem{Chan:1996}
H.~Chan, J.~Faridani and S.~T.~Tsou,
Phys.\ Rev.\  {\bf D52} (1995) 6134,
[hep-th/9503106];\\
H.~Chan, J.~Faridani and S.~Tsou,
Phys.\ Rev.\  {\bf D53} (1996) 7293,
[hep-th/9512173];\\
H.~Chan and S.~T.~Tsou,
Int.\ J.\ Mod.\ Phys.\  {\bf A14} (1999) 2139,
[hep-th/9904102].

\bibitem{M5}
P.~Pasti, D.~Sorokin and M.~Tonin,
Phys.\ Lett.\  {\bf B398} (1997) 41,
[hep-th/9701037].

\bibitem{fks}
 L.~De Fosse, P.~Koerber and A.~Sevrin,
[hep-th/0103015].

\bibitem{Vinogradov}
A.M. Vinogradov, {\em Sov. Math. Dokl.\/} {\bf 18} (1977) 1200;\\
A.M. Vinogradov, {\em Sov. Math. Dokl.\/} {\bf 19} (1978) 144.

\bibitem{Newton}
H.M. Edwards, {\em Galois theory} (Springer-Verlag, 1984), p. 6.

\bibitem{Dieu}  
J. Dieudonn\'e, {\em Fondements de l'analyse moderne} (Gauthiers-Villars, 1967), p. 263.

\end{thebibliography}
\end{document}